# Improved Blind Channel Estimation Performance by Nearby Channel Estimation

Jia-Chyi Wu and Zhen-Wei Kao

Department of Communications, Control and Navigation Engineering,

National Taiwan Ocean University

Email: jcwu@ntou.edu.tw

## Abstract

We have investigated the impact of adjacent channel estimation information on blind channel estimation performances. To obtain channel information for further data transmission, the Blind channel estimation needs no pilot signal in advance is a considerable algorithm. To improve performance of the blind channel estimation schemes, we have employed the channel estimation with pilot signals by adjacent users; the acquired estimation channel information of the adjacent channel is applied as an initial state to the blind channel estimation of the target user. We have considered the subspace blind channel estimation scheme in this study. The estimated adjacent channel information is supported as an initial state of the autocorrelation matrix in the subspace estimation method, thereby providing information to reduce the channel estimation error. From our simulation study, we have found that, when the correlation between two nearby channels is high, the forgetting factor in the subspace estimation method can be increased, so that the weight of the initial state of the matrix is increased, and a more accurate channel state is estimated. If the correlation coefficient is lowered in two adjacent channels, the forgetting factor can be appropriately adjusted to rise the weight of the direct received signal from the objective channel to avoid errors increased.

Keywords: Pilot Signal, Subspace Blind Channel Estimation, Nearby Channel Estimation.

## Introduction

Signal transmitting over the wireless communication channel is effected by the physical characteristics of the electromagnetic wave propagation and the obstacles of the communication paths where, diffraction, refraction, and reflection multipath propagation interference happened at the receiver to destroy the signal quality. Most of the signal interferences can be eliminated accordingly by applicable equalization process, but depends on a good channel estimation algorithm to get more real and accurate channel information.

The channel estimation technology can be roughly divided into three categories. The first is the index estimation method by transmitting the pilot signal first. The receiver may estimate the channel information from the received pilots by observing the variation of the received known signals after passing through the channel. Since the pilot signal is an extra data frame apart from the demanded data streams, there is a disadvantage of bandwidth efficacy. The second type of estimation method relies on the demand data signal received to make the channel estimation. The received signal streams are not well known signal information at receiver to serve as an estimation references, this is so called a blind estimation method. This kind of schemes may save bandwidth requests to increase data transmission rates, but the system is more complicated. The third kind of methods attempts combining the advantages from the above two estimation categories to develop a better channel estimation scheme. In this paper, we have combined the blind estimation method with the index estimation results of the nearby channel to improve blind channel estimation performance. The estimated nearby channel information is supported as an initial state of the autocorrelation matrix in the subspace estimation method, thereby providing information to reduce the channel estimation error.

## CHANNEL ESTIMATION SYSTEM

Signal transmission over the wireless communication channel is effected by the physical characteristics of the electromagnetic wave propagation and the obstacles of the communication paths where, diffraction, refraction, and reflection multipath propagation interference happened at the receiver to destroy the signal quality. Most of these interferences can be reduced by the receiver design and applicable equalizer, but depends on a good channel estimation algorithm to get more real and accurate channel information (Figure 1).

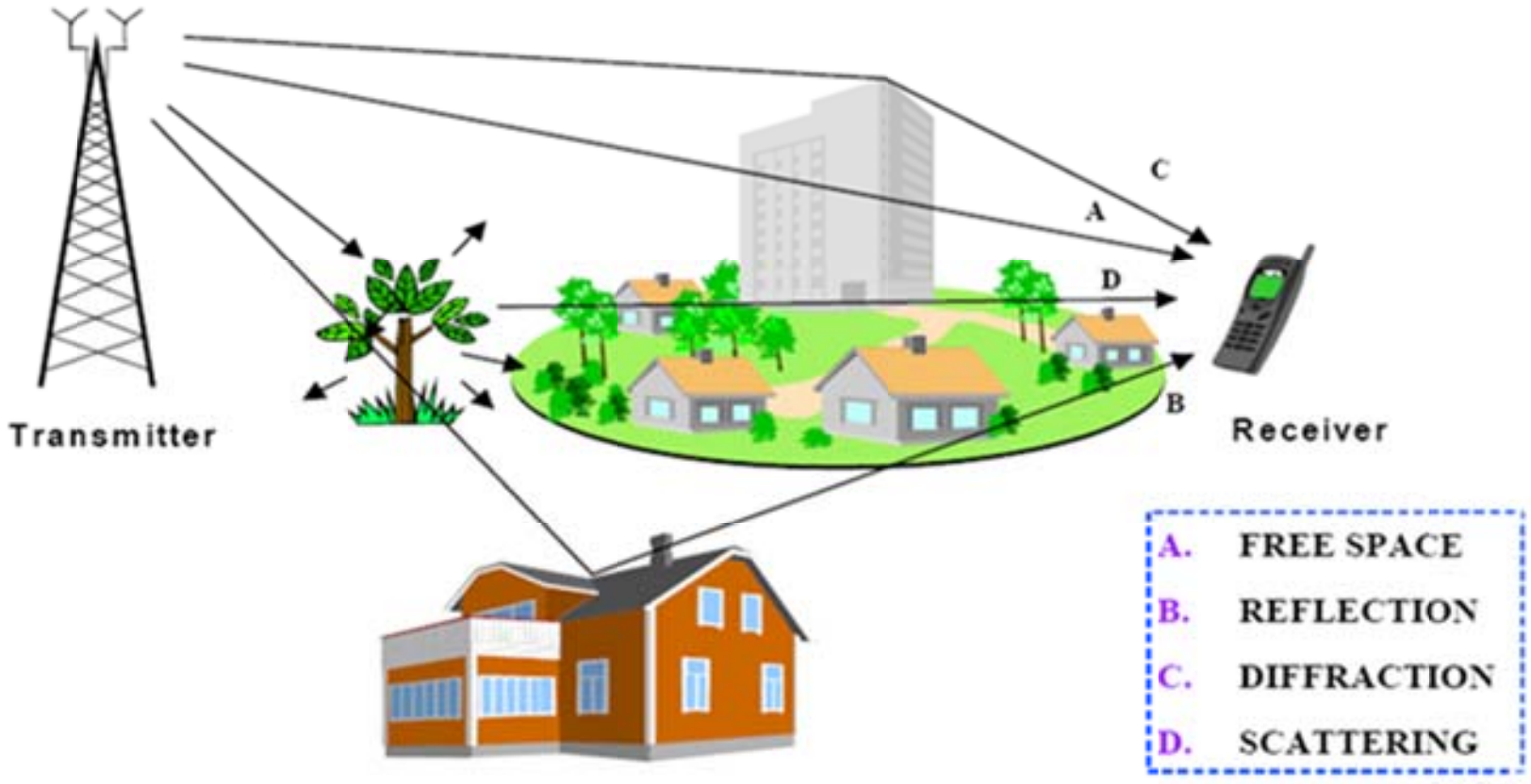

Figure 1. Data transmission channel

Conventionally, transmitting training sequences or pilots signal known at the receiver as a reference for estimation may reach the prerequisite. By observing the changes and variation of these known signals after passing through the transmitted channel leads to a better estimated channel information. However, transmission of these known pilot signals consumes bandwidth resources yet inefficient for a fast varying channel conditions. Blind channel estimation techniques need no further known signals in advance been proposed to improve the bandwidth requirement.

## A. Blind Channel Estimation

The blind channel estimation algorithm estimates the channel impulse response **H**, from the received signal **y** (which contains noise).

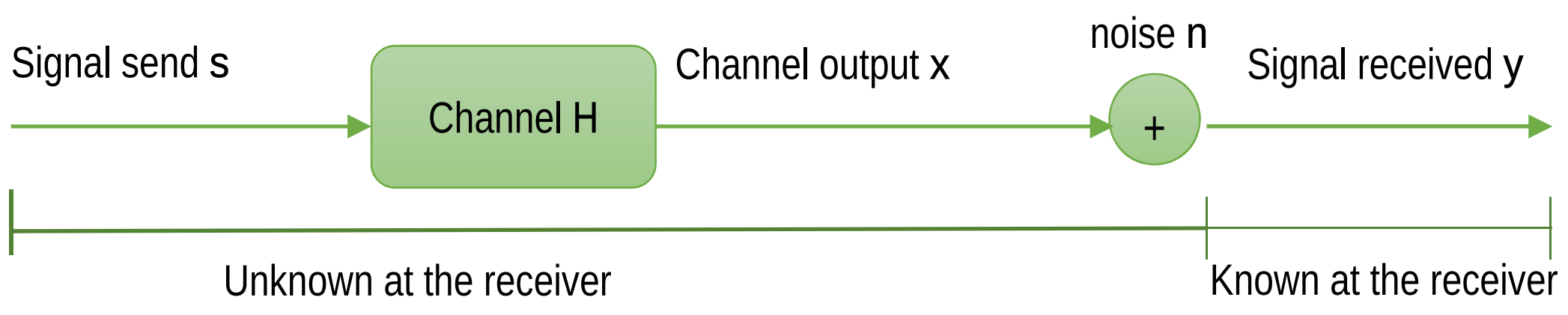

Figure 2. Channel estimation algorithm concept

There are two categories of the blind estimation algorithms: one needs the probability statistical characteristics of the transmitted signals, which is called statistical methods; the other one is classified as deterministic methods if the probability features of the transmitted signals are not required. In real states, the receiver does not make statistical assumptions about the transmitted signals, therefore the deterministic methods is more realistic.

## B. Estimation System and Channel Model

We have considered the subspace method as the estimation method which is combined with Single-Input Multi-Output (SIMO) technology and over-sampling. The autocorrelation matrix of the received signal can be decomposed into signal subspace and noise subspace [3, 4]. Then, B. Muquet and M. de Courville [5, 6] applied to CP-OFDM systems with the cyclic prefix (CP) scheme, but no need of the SIMO and oversampling techniques. Considering the OFDM system, the received signal $\boldsymbol{r}_P(k)$ is segmented with its CP length G = 1⁄4 M, M is the number of subcarriers:

$$\boldsymbol{r}_{\mathrm{p}}(k)=[\,\boldsymbol{r}_{s1}(k)^T,\ \ldots,\ \boldsymbol{r}_{s5}(k)^T\,]^T \tag{1}$$

The transmission signal $\boldsymbol{s}_P(k)$ and noise $\boldsymbol{n}_P(k)$ have been divided into 5 segments:

$$\boldsymbol{s}_{\mathrm{p}}(k)=[\,\boldsymbol{s}_{s1}(k)^T,\ \ldots,\ \boldsymbol{s}_{s5}(k)^T\,]^T \tag{2}$$

$$\boldsymbol{n}_{\mathrm{p}}(k)=[\,\boldsymbol{n}_{s1}(k)^T,\ \ldots,\ \boldsymbol{n}_{s5}(k)^T\,]^T \tag{3}$$

The received signal $\boldsymbol{r}_P(k)$ can be obtained from $\boldsymbol{s}_P(k)$ by the following equation ($\boldsymbol{H}_i$ is the channel vector coefficient):

$$\boldsymbol{r}_{\mathrm{p}}(k) = \boldsymbol{H_0}\boldsymbol{s}_{\mathrm{p}}(k) + \boldsymbol{H_1}\boldsymbol{s}_{\mathrm{p}}(k-1) + \boldsymbol{n}_{\mathrm{p}}(k) \tag{4}$$

The first column vector is $[h_0, \ldots, h_{P-1}]^T$ and the first row vector is $[h_0, 0, \ldots, 0]$ of the P×P Toeplitz matrix $\boldsymbol{H}_0$. The first column vector is $[0, \ldots, 0]^T$ and the first row vector is $[0, h_{P-1}, \ldots, h_1]$ of the P×P Toeplitz matrix $\boldsymbol{H}_0$. We have assumed that the channel vector length is $L + 1$ for $\mathbf{h} = [h_0, h_1, \ldots, h_L]^T$, where $L < G < P$), when $j > L$ then $h_j = 0$, i.e.:

$$\mathbf{h} = \left[h_0, h_1, \ldots, h_P\right]^T, \; h_j = 0, \; for\, j > L \tag{5}$$

We may rewrite Eq. (4) to be:

$$\boldsymbol{r}_{\mathrm{p}}(k) = \begin{bmatrix} \boldsymbol{C_1} & 0 & 0 & 0 & \boldsymbol{C_0} \\ 0 & \boldsymbol{C_0} & 0 & 0 & \boldsymbol{C_1} \\ 0 & \boldsymbol{C_1} & \boldsymbol{C_0} & 0 & 0 \\ 0 & 0 & \boldsymbol{C_1} & \boldsymbol{C_0} & 0 \\ 0 & 0 & 0 & \boldsymbol{C_1} & \boldsymbol{C_0} \end{bmatrix} \begin{bmatrix} \boldsymbol{s}_{s4}(k-1) \\ \boldsymbol{s}_{s1}(k) \\ \boldsymbol{s}_{s2}(k) \\ \boldsymbol{s}_{s3}(k) \\ \boldsymbol{s}_{s4}(k) \end{bmatrix} + \begin{bmatrix} \boldsymbol{n}_{s1}(k) \\ \boldsymbol{n}_{s2}(k) \\ \boldsymbol{n}_{s3}(k) \\ \boldsymbol{n}_{s4}(k) \\ \boldsymbol{n}_{s5}(k) \end{bmatrix} \tag{6}$$

The matrices $\boldsymbol{C}_0$ and $\boldsymbol{C}_1$ are $G \times G$ Toeplitz matrices. The received signal $\bar{\boldsymbol{r}}(k)$ of size $(2M+G)\times 1$ is formed by two consecutive received signals $\boldsymbol{r}_P(k)$ and $\boldsymbol{r}_P(k\text{-}1)$:

$$\bar{\boldsymbol{r}}(k) = [\boldsymbol{r}_{s2}(k-1)^T, \ldots, \boldsymbol{r}_{s5}(k-1)^T, \boldsymbol{r}_{s1}(k)^T, \ldots, \boldsymbol{r}_{s5}(k)^T]^T \tag{7}$$

The similar format for $\bar{\boldsymbol{s}}(k)$ and $\bar{\boldsymbol{n}}(k)$ can be found. Threfore, the relationship between$\bar{\boldsymbol{s}}(k)$ and $\bar{\boldsymbol{r}}(k)$ can be joined by $(2M+G)\times 2M$ matrix $\boldsymbol{H}(\boldsymbol{h})$:

$$\bar{\boldsymbol{r}}(k) = \boldsymbol{H}(\boldsymbol{h})\bar{\boldsymbol{s}}(k) + \bar{\boldsymbol{n}}(k) \tag{8}$$

$$\boldsymbol{H}(\boldsymbol{h}) = \begin{bmatrix} \boldsymbol{C_0} & \boldsymbol{0} & \boldsymbol{0} & \boldsymbol{C_1} & \boldsymbol{0} & \boldsymbol{0} & \boldsymbol{0} & \boldsymbol{0} \\ \boldsymbol{C_1} & \boldsymbol{C_0} & \boldsymbol{0} & \boldsymbol{0} & \boldsymbol{0} & \boldsymbol{0} & \boldsymbol{0} & \boldsymbol{0} \\ \boldsymbol{0} & \boldsymbol{C_1} & \boldsymbol{C_0} & \boldsymbol{0} & \boldsymbol{0} & \boldsymbol{0} & \boldsymbol{0} & \boldsymbol{0} \\ \boldsymbol{0} & \boldsymbol{0} & \boldsymbol{C_1} & \boldsymbol{C_0} & \boldsymbol{0} & \boldsymbol{0} & \boldsymbol{0} & \boldsymbol{0} \\ \boldsymbol{0} & \boldsymbol{0} & \boldsymbol{0} & \boldsymbol{C_1} & \boldsymbol{0} & \boldsymbol{0} & \boldsymbol{0} & \boldsymbol{C_0} \\ \boldsymbol{0} & \boldsymbol{0} & \boldsymbol{0} & \boldsymbol{0} & \boldsymbol{C_0} & \boldsymbol{0} & \boldsymbol{0} & \boldsymbol{C_1} \\ \boldsymbol{0} & \boldsymbol{0} & \boldsymbol{0} & \boldsymbol{0} & \boldsymbol{C_1} & \boldsymbol{C_0} & \boldsymbol{0} & \boldsymbol{0} \\ \boldsymbol{0} & \boldsymbol{0} & \boldsymbol{0} & \boldsymbol{0} & \boldsymbol{0} & \boldsymbol{C_1} & \boldsymbol{C_0} & \boldsymbol{0} \\ \boldsymbol{0} & \boldsymbol{0} & \boldsymbol{0} & \boldsymbol{0} & \boldsymbol{0} & \boldsymbol{0} & \boldsymbol{C_1} & \boldsymbol{C_0} \end{bmatrix} \tag{9}$$

We are able to calculate the autocorrelation matrix from $\bar{\boldsymbol{r}}(k)$ by $\boldsymbol{R}_{\bar{\boldsymbol{r}}\bar{\boldsymbol{r}}} = E\left[\,\bar{\boldsymbol{r}}(k)\bar{\boldsymbol{r}}(k)^H\,\right]$,

$$\begin{aligned} \boldsymbol{R}_{\bar{\boldsymbol{r}}\bar{\boldsymbol{r}}} &= E[\,\bar{\boldsymbol{r}}(k)\bar{\boldsymbol{r}}(k)^H\,] \\ &= E\left[(\boldsymbol{H}(\boldsymbol{h})\bar{\boldsymbol{s}}(k) + \bar{\boldsymbol{n}}(k))(\boldsymbol{H}(\boldsymbol{h})\bar{\boldsymbol{s}}(k) + \bar{\boldsymbol{n}}(k))^H\right] \\ &= E[\boldsymbol{H}(\boldsymbol{h})\bar{\boldsymbol{s}}(k)\bar{\boldsymbol{s}}(k)^H\boldsymbol{H}(\boldsymbol{h})^H] + E[\,\bar{\boldsymbol{n}}(k)\bar{\boldsymbol{n}}(k)^H\,] \\ &= \boldsymbol{H}(\boldsymbol{h})\boldsymbol{R}_{\bar{\boldsymbol{s}}\bar{\boldsymbol{s}}}\boldsymbol{H}(\boldsymbol{h})^H + \sigma^2 I \end{aligned} \tag{10}$$

After applying singular value decomposition (SVD), we can find the orthogonal relationship of the channel matrix $\boldsymbol{H}(\boldsymbol{h})$ and matrix $\boldsymbol{G}_j$ formed by eigenvector $\boldsymbol{u}_j$ of the autocorrelation matrix:

$$\boldsymbol{h}^H \boldsymbol{G}_j = \boldsymbol{0}\,, for\ 0 \le j \le G-1 \tag{11}$$

After we find the matrix $\boldsymbol{G}_j$, the channel coefficient vector $\boldsymbol{h}$ can be estimated by Eq. (11).

From Eq. (11), we may obtain the channel information by channel coefficient vector $\boldsymbol{h}$, which is orthogonal to the matrix $\boldsymbol{G}_j$. In fact, there exist an uncertainty about magnitude and phase ambiguity between $\widehat{\boldsymbol{h}}$, the estimated channel coefficient vector and the actual channel coefficient $\boldsymbol{h}$:

$$\widehat{\boldsymbol{h}} = \alpha \boldsymbol{h} \tag{12}$$

where $\alpha$ is a complex scalar. In order to reduce the impact of this uncertainty, some pilot signals should be inserted at the time of transmission.

### C. Blind Channel Estimation - subspace method with deterministic condition

By the subspace approach, we first assume a SIMO channel with P outputs, the output $\vec{\boldsymbol{y}}_q(k)$ (with $q$ symbols) can be obtained by,

$$\vec{\boldsymbol{y}}_q(k) = \vec{\boldsymbol{x}}_q + \vec{\boldsymbol{n}}_q, \text{ with } \vec{\boldsymbol{x}}_q(k) = F_q(\vec{\boldsymbol{h}})\, \vec{\boldsymbol{s}}_{q+L}(k) \tag{13}$$

$$\vec{\boldsymbol{n}}_q(k) = [\boldsymbol{n}_k^T, \ldots, \boldsymbol{n}_{k-q+1}^T]^T \tag{14}$$

$$F_q(\vec{\boldsymbol{h}}) \triangleq \begin{pmatrix} \boldsymbol{h}0 & \ldots & \boldsymbol{h}L & & \\ & \ddots & & \ddots & \\ & & \boldsymbol{h}0 & \ldots & \boldsymbol{h}L \end{pmatrix} \tag{15}$$

Therefore, we may rewrite (13) as $\boldsymbol{Y}_q = \boldsymbol{H}_q \boldsymbol{S}_q + \boldsymbol{N}_q$, the autocorrelation matrix, $\boldsymbol{R_Y}$ is calculated by $\boldsymbol{Y}_q$,

$$\begin{aligned} \boldsymbol{R_Y} &\triangleq E(\boldsymbol{Y}_q \boldsymbol{Y}_q^H) \quad dim.\ qP \times qP \\ &= E\left((\boldsymbol{H}_q \boldsymbol{S}_q + \boldsymbol{N}_q)(\boldsymbol{H}_q \boldsymbol{S}_q + \boldsymbol{N}_q)^H\right) \\ &= E(\boldsymbol{H}_q \boldsymbol{S}_q \boldsymbol{S}_q^H \boldsymbol{H}_q^H) + E(\boldsymbol{N}_q \boldsymbol{N}_q^H) \\ &= \boldsymbol{H}_q \boldsymbol{R_S} \boldsymbol{H}_q^H + \sigma^2 \boldsymbol{I} \end{aligned} \tag{16}$$

The autocorrelation matrix of the transmitted signal is $\boldsymbol{R_S}$ ($dim.(L+q)\times(L+q)$), and $\boldsymbol{N}_q$ is zero mean, variance $\sigma^2$ additive white Gaussian noise, where its autocorrelation matrix $\boldsymbol{R_N}$ ($dim.(Lq)\times(Lq)$) expressed as $\sigma^2\boldsymbol{I}$. $\boldsymbol{R_Y}$ can be eigen-decomposite, i.e.

$$\boldsymbol{R_Y} = \sum_{i=0}^{P(q-L)-1} \lambda_i \boldsymbol{v}_i \boldsymbol{v}_i^H \tag{17}$$

where $\lambda_i$ are the eigenvalues, and $\boldsymbol{v}_i$ are the corresponding eigenvectors.

Re-arrange the eigenvalues from large to small, $\lambda_0 > \lambda_1 > \ldots > \lambda_{P(q-L)-1}$, since $\boldsymbol{H}_q$ and $\boldsymbol{R}_S$ are full-column rank, the rank of $\boldsymbol{H}_q\boldsymbol{R}_S\boldsymbol{H}_q^H$ is $L+q$. Therefore, the eigenvalues possess the following characteristics:

$$\lambda_i > \sigma^2,\ i = 0, 1, \ldots, L+q-1$$

$$\lambda_i = \sigma^2,\ i = L+q, L+q+1, \ldots, Pq-1 \tag{18}$$

When $\lambda_i = \sigma^2$ the corresponding eigenvectors, $\boldsymbol{v}_i$ may yield ($\boldsymbol{A}\vec{\boldsymbol{x}} = \lambda\vec{\boldsymbol{x}}$):

$$\boldsymbol{R_Y}\boldsymbol{v}_i = \sigma^2\boldsymbol{v}_i,\ i = L+q, L+q+1, \ldots, Pq-1 \tag{19}$$

$$\left(\boldsymbol{H}_q\boldsymbol{R}_s\boldsymbol{H}_q^H + \sigma^2\boldsymbol{I}\right)\boldsymbol{v}_i = \sigma^2\boldsymbol{v}_i \xrightarrow{\text{yields}} \boldsymbol{H}_q\boldsymbol{R}_s\boldsymbol{H}_q^H\boldsymbol{v}_i + \sigma^2\boldsymbol{I}\boldsymbol{v}_i - \sigma^2\boldsymbol{v}_i = 0$$

Therefore, $\boldsymbol{H}_q\boldsymbol{R}_S\boldsymbol{H}_q^H = 0$ where, $\boldsymbol{H}_q$ and $\boldsymbol{R}_S$ are full-column rank, the unknown multi-channel system matrix $\boldsymbol{H}_q^H$ are orthogonal to the eigenvectors for the corresponding eigenvalues $\sigma^2$. There are $P_q$ eigenvectors of the autocorrelation matrix $\boldsymbol{R_Y}$, can be divided into two spans, one is the signal subspace, and the other one is noise subspace. For eigenvalues $\lambda_i > \sigma^2$, $i = 0, 1, \ldots, L+q-1$, corresponding eigenvectors are $\boldsymbol{s}_i$, and for $\lambda_i = \sigma^2$, $i = L+q, L+q+1, \ldots, Pq-1$, corresponding eigenvectors are $\boldsymbol{g}_i$, we may define the two subspaces:

Signal subspace: $\boldsymbol{S}=[\boldsymbol{s}_0 \ \cdots \ \boldsymbol{s}_{L+q-1}]$

Noice subspace: $\boldsymbol{G}=[\boldsymbol{g}_0 \ \cdots \ \boldsymbol{g}_{\boldsymbol{Pq-L-q-1}}]$

Suppose the rank of $\boldsymbol{Y}_q$ is $P_q$ and $P_q > L+q$, since the rank of $\boldsymbol{H}_q\boldsymbol{R_S}\boldsymbol{H}_q^H$ is $L+q$, which means that the signal subspace is generated by $L+q$ eigenvectors, so the noise subspace is generated by the left $P_q-(L+q)$ eigenvectors. The rank of $\boldsymbol{H}_q\boldsymbol{R_S}\boldsymbol{H}_q^H$ must be less than $\boldsymbol{Y}_q$ for the estimation method applicable to find out two subspaces separately.

The noise eigenvectors is orthogonal to the channel matrix $\boldsymbol{H}_q$, where the noise subspace generated by these eigenvectors is also orthogonal to $\boldsymbol{H}_q$. Since the signal subspace can be generated by the column vector of $\boldsymbol{H}_q$, the orthogonal affiliation between the noise subspace and the signal subspace can be achieved. Therefore, the channel can be solved by the equation $\boldsymbol{H}_q^H \boldsymbol{g}_i = 0$ in conjunction with the least squares sense. The accurate initial state $\boldsymbol{R}_Y^P$ can be effectively reduced by the index estimation to effectively reduce the complex scalar $\alpha$.

## NEARBY CHANNEL CONCEPT

We first define the nearby channel concept for transmission with the main channel and the idle unused channel as a nearby channel (Figure 3). Channel condition information from nearby channel can be accessed to help the data transmission in the main channel. Nearby channel condition information is utilized to improve the blind subspace channel estimation in the main channel.

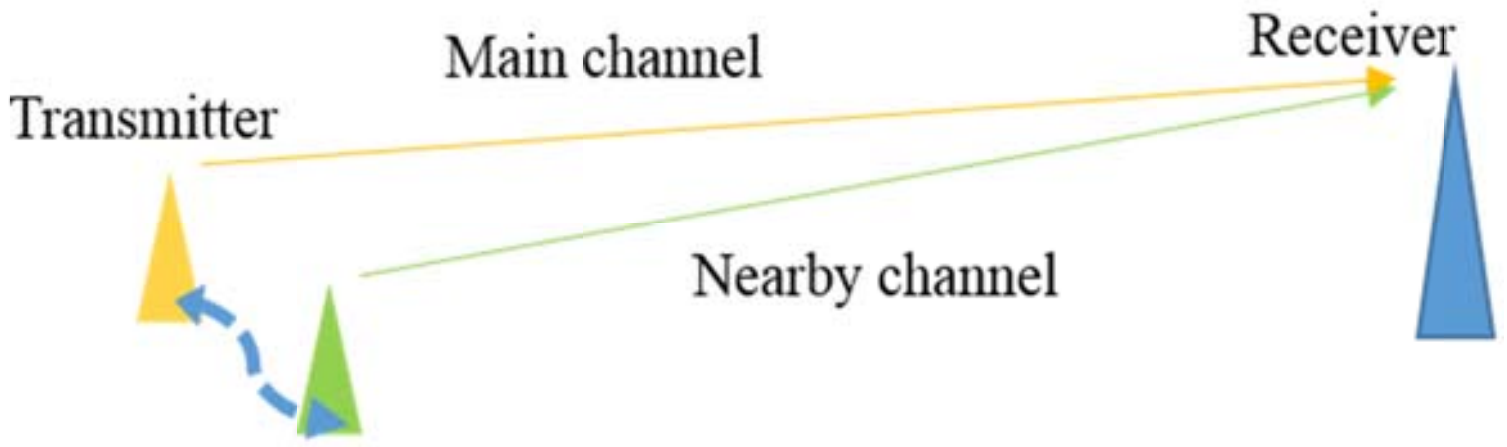

Figure 3. Nearby channel concept

### A. Simulation Setup

We have learned that channel estimation with pilot sequence achieve better accuracy than the blind estimation algorithm. Assume two channels A and B are adjacent as shown in Figure 4, channel B is idle transmitting; we may apply channel estimation pilot sequence constantly performed on this channel, where channel A is in data transmission mode with blind channel estimation.

Experimental analysis is performed at nearby distance (3 to 5 meters) to test the channel condition likeness; consequently, the estimation results of channel B are applied as initial states for blind channel estimation at channel A.

Channel estimation accuracy is improved for data transmission on channel A without excess bandwidth waste.

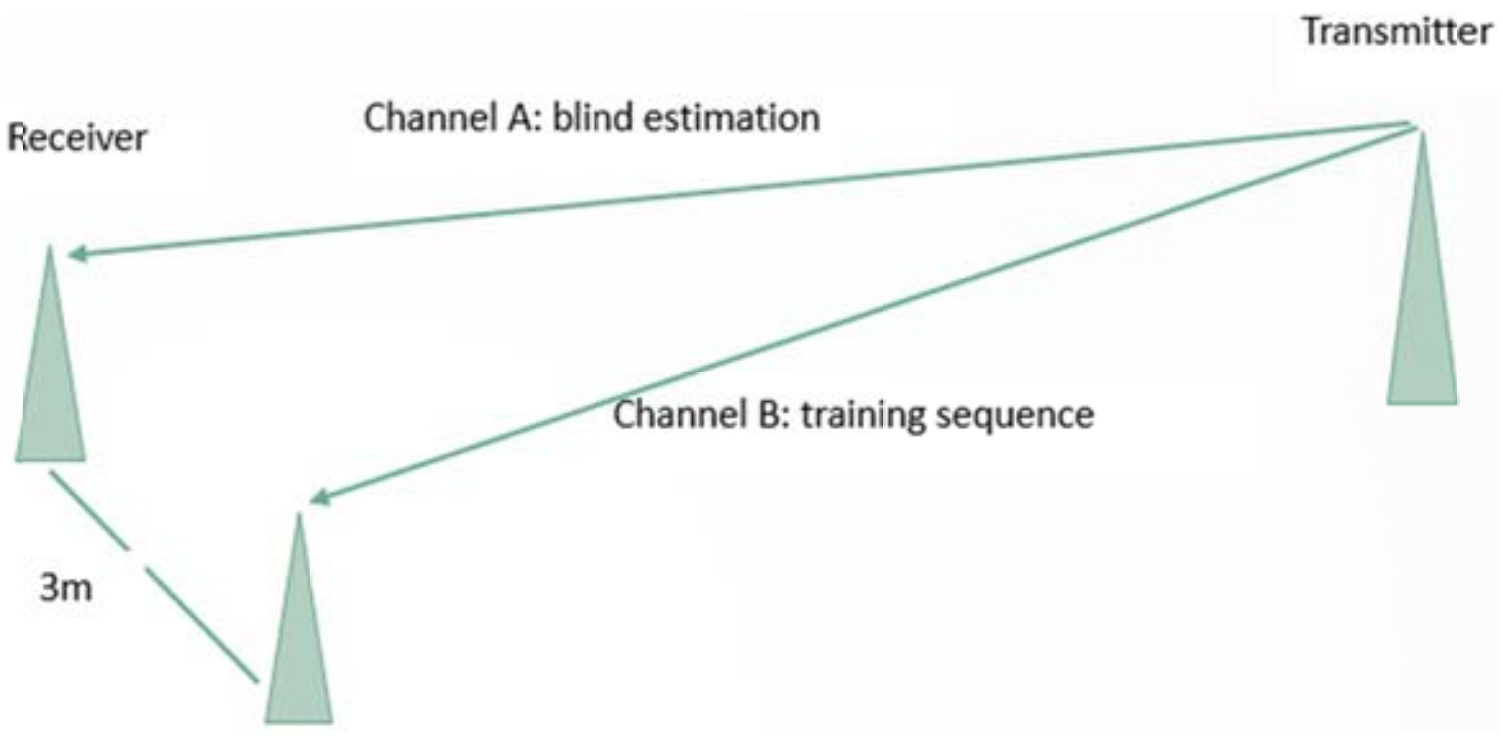

Figure 4. Nearby channel concept simulation setup.

## B. Nearby Channel Condition Measurements

Real-time received signal power profile measurements for nearby channels [7], at distance around 3 to 5 meters are experimental tested, the channel power profile conditions likeness are measured as shown in Figures 5-7.

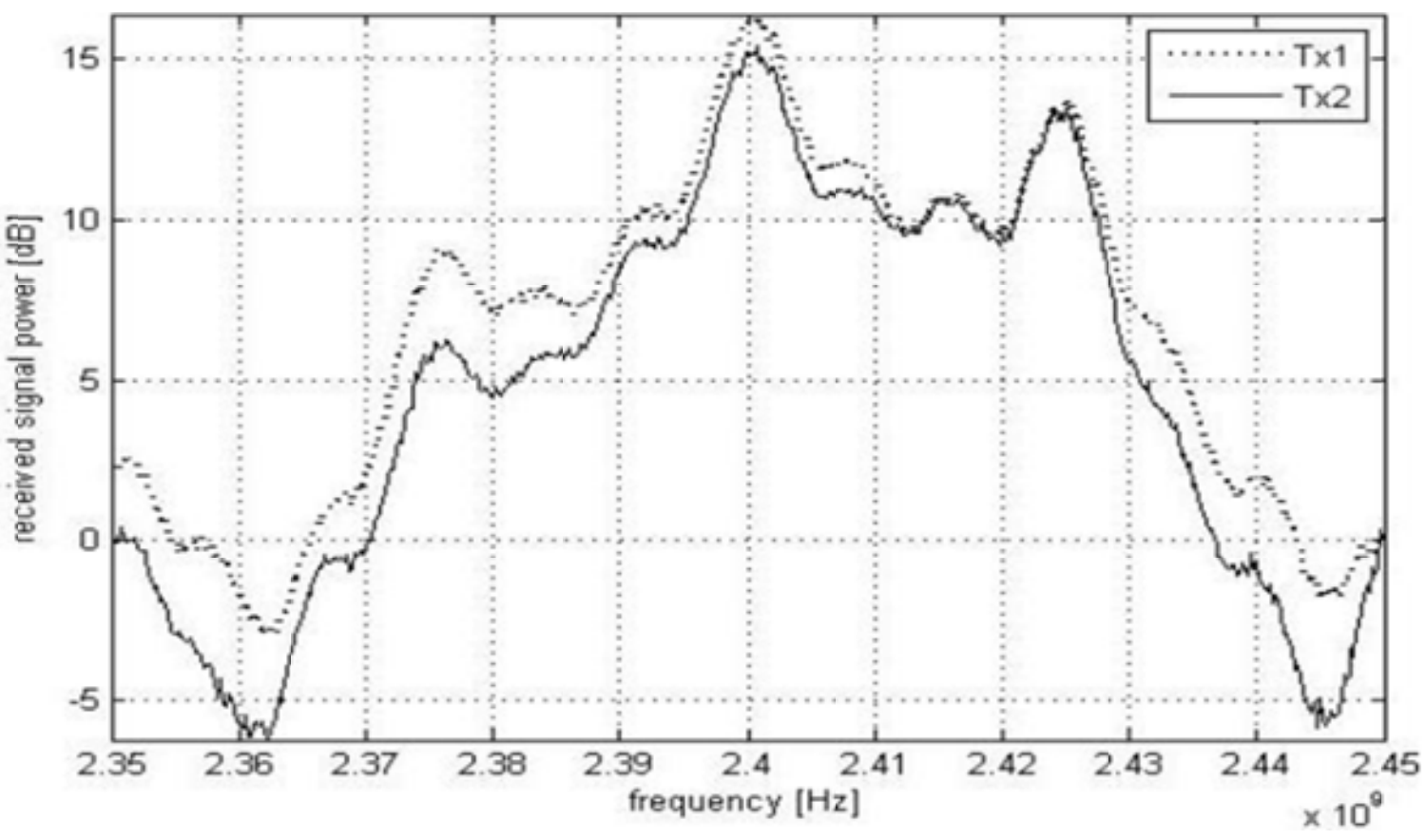

Figure 5. 3-meters apart (correlation coefficient 0.98)

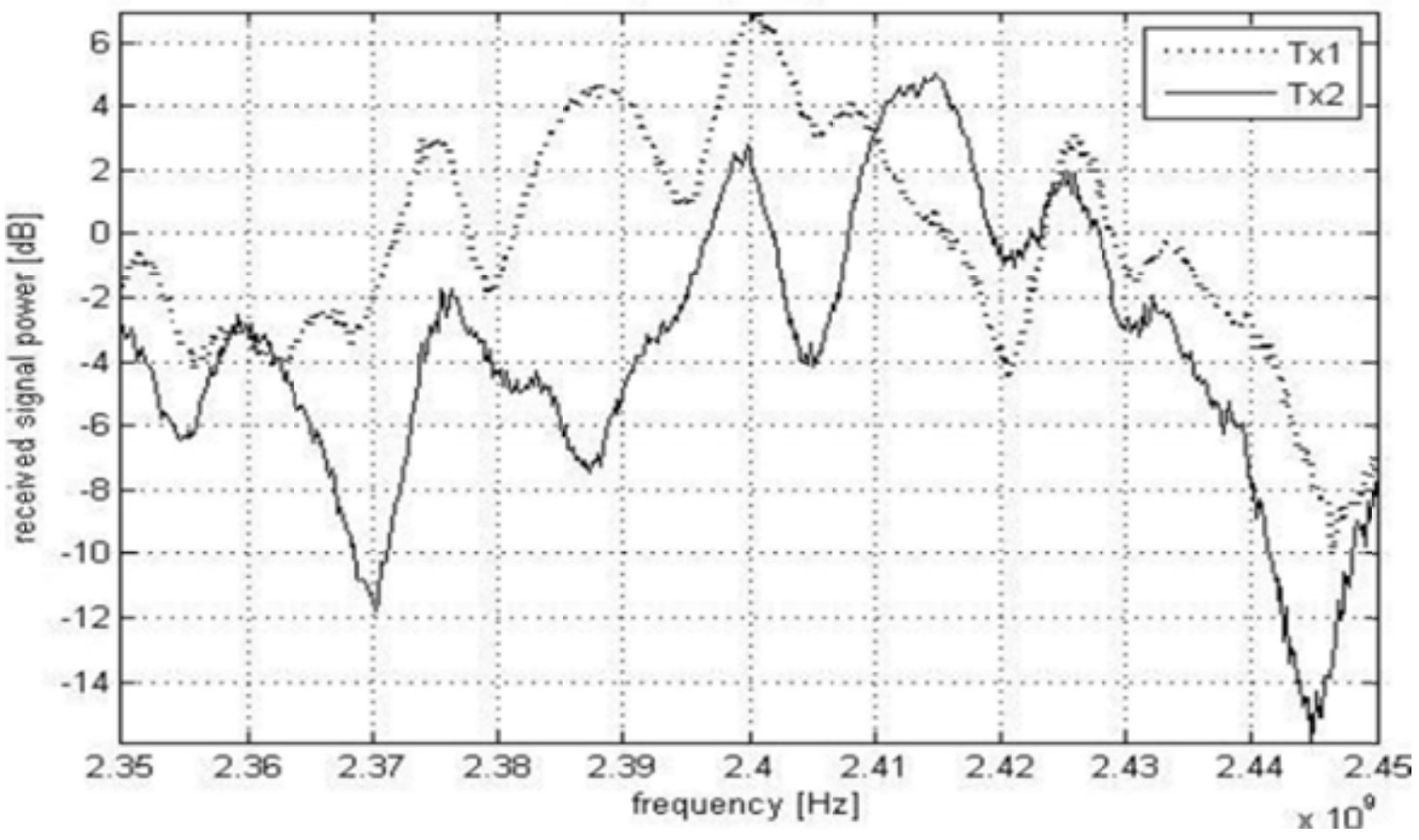

Figure 6. 4-meters apart (correlation coefficient 0.68)

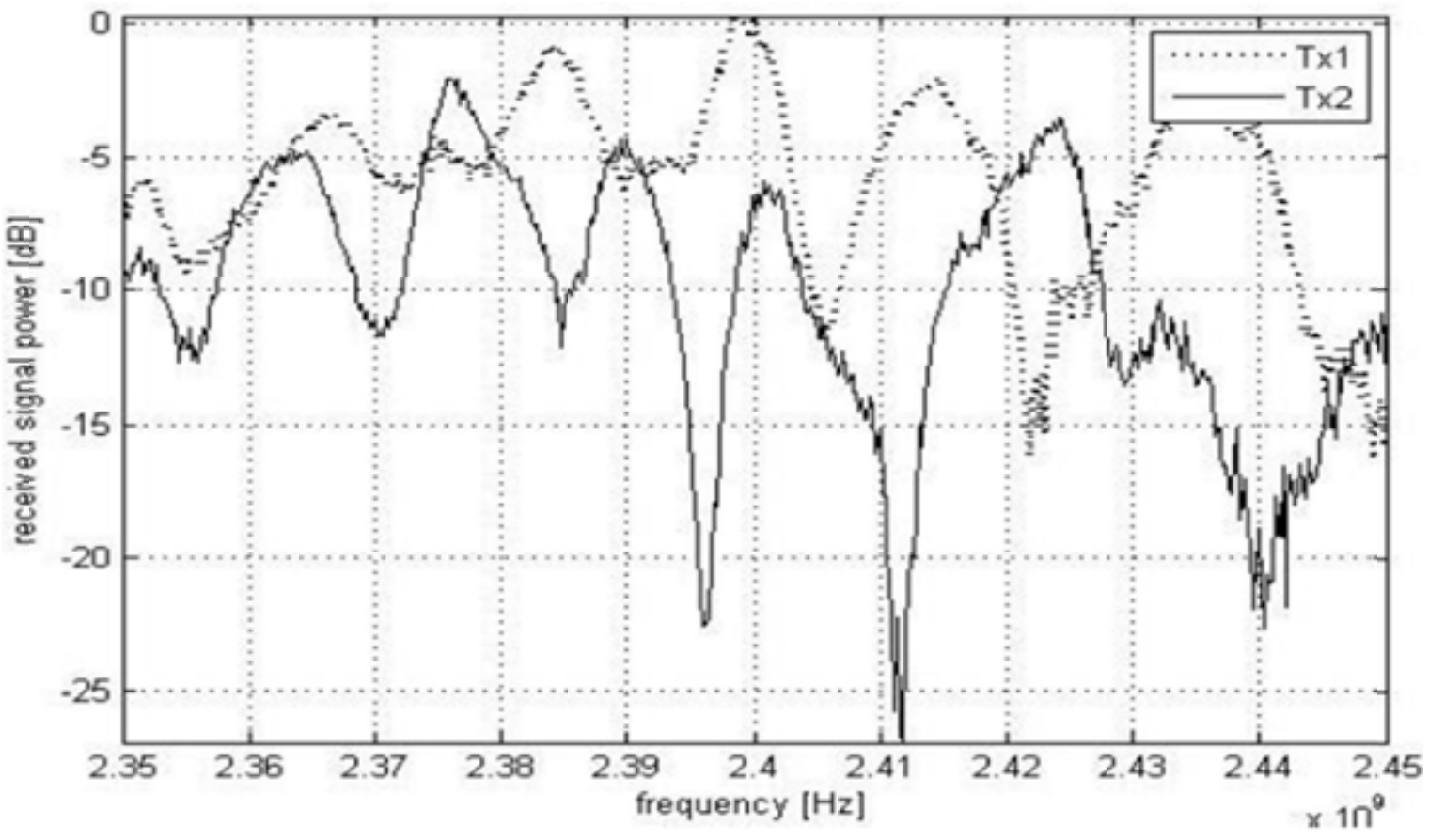

Figure 7. 5-meters apart (correlation coefficient 0.11)

Based on the received power profile correlation coefficients between two nearby adjacent channels by real-time measurements results, we can perform the numerical simulations for nearby channel condition by mathematical models, as shown in Figures 8-10.

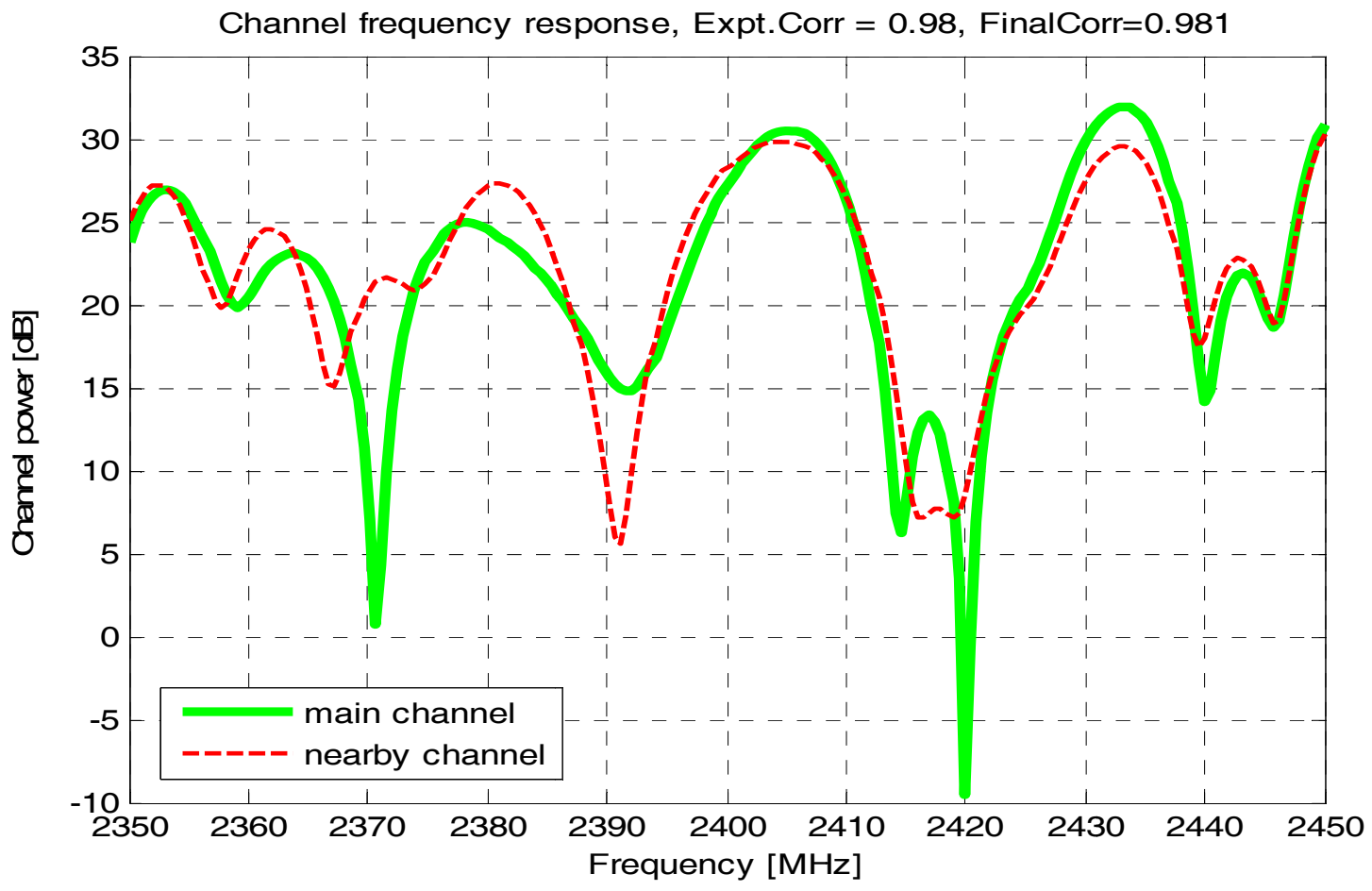

Figure 8. Expected correlation coefficient 0.98

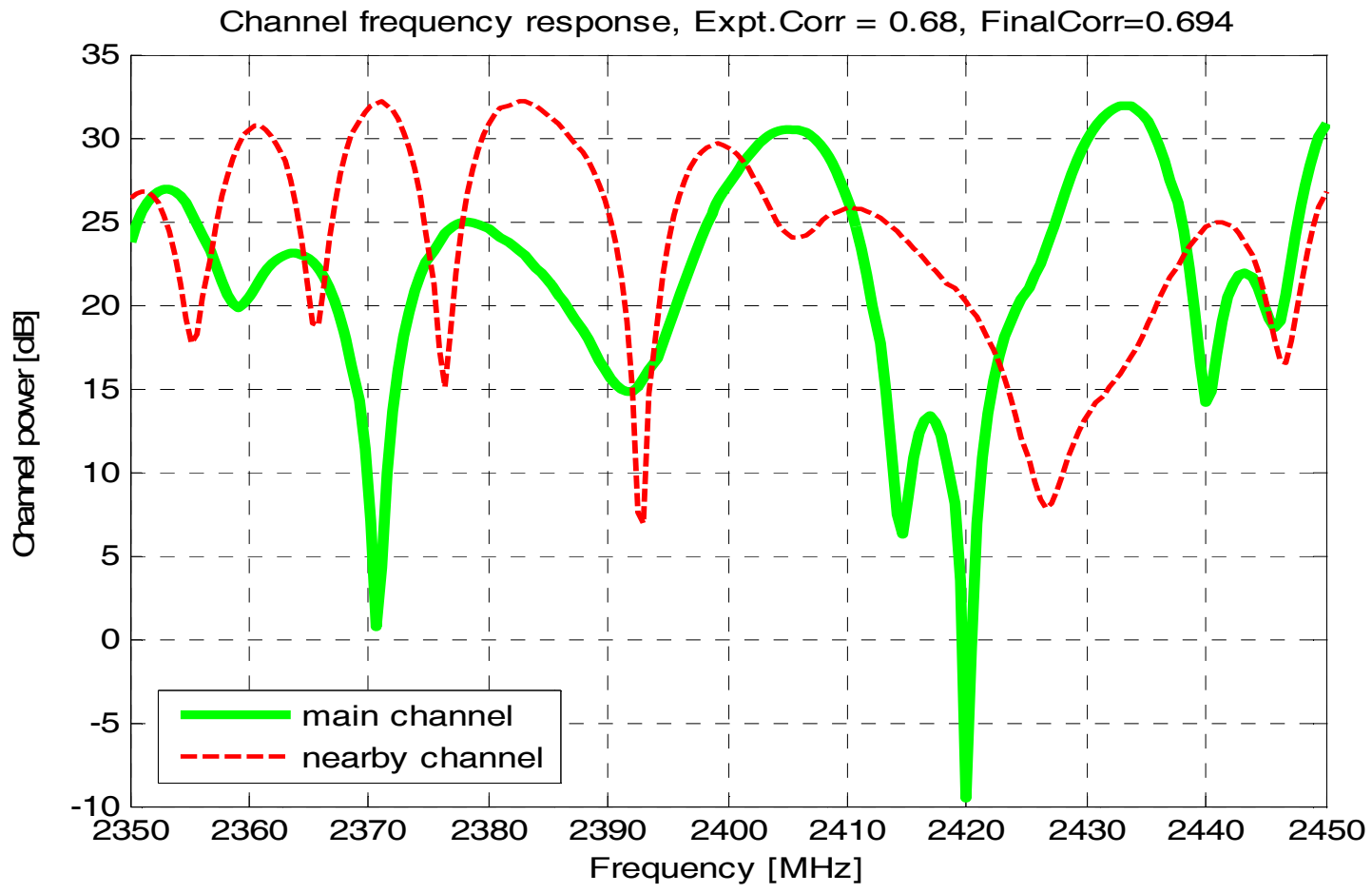

Figure 9. Expected correlation coefficient 0.68

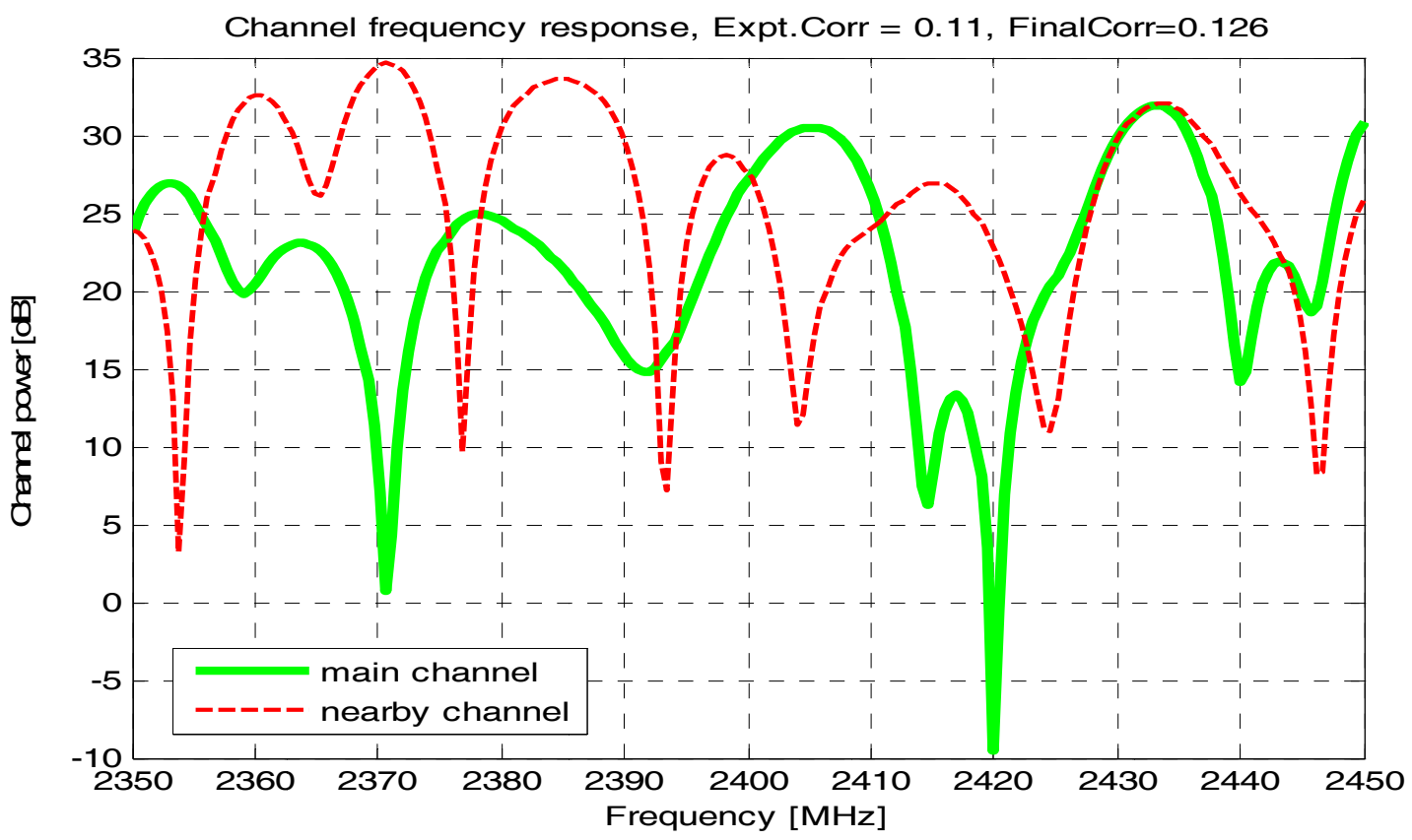

Figure 10. Expected correlation coefficient 0.11

## SIMULATION RESULTS ANALYSIS

We are able to apply nearby channel condition information to blind channel estimation based on the previous channel measurement results. The system simulation setup diagram is shown in Figure 11.

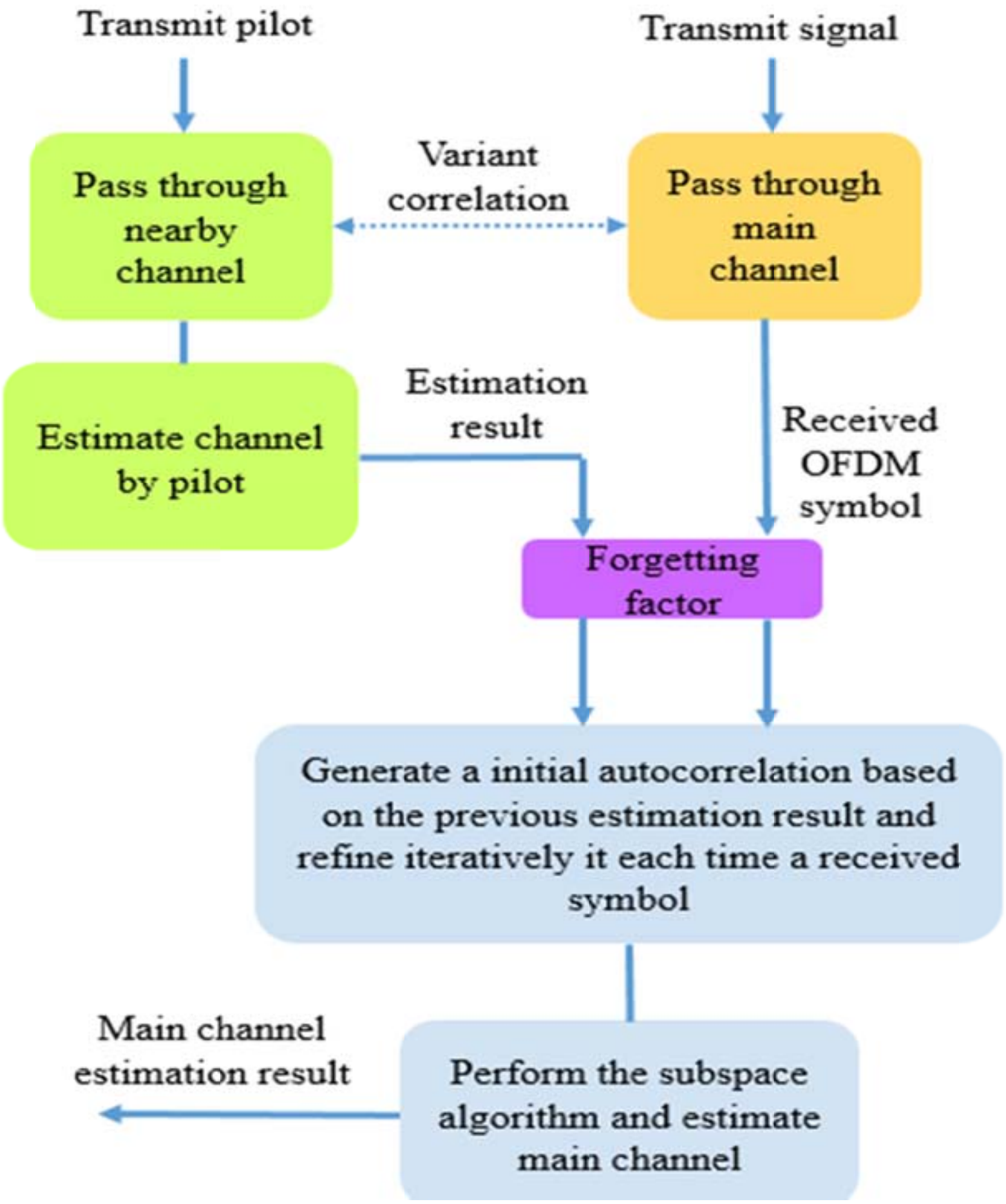

Figure 11. Systematic Block Diagram

Forgetting factor represents the weight of the previous state and the next state in the process of updating the autocorrelation matrix. If the forgetting factor with higher value, which means the present channel state is more effect by previous state. Since the initial state information is provided by the estimation result from the nearby channel, these two nearby channels being similar or not will greatly affect the final estimation result. If the forgetting factor is with small value, the influence of the adjacent channel information will be reduced, and the weight of each received symbol pair estimation will be increased.

Simulations are based on Cyclic Prefix OFDM (CP-OFDM) with QPSK modulation technology. Channel estimation performances are varied by the nearby distances of the two transmitting antennas in 3~5 meters. System simulation parameters setting are as follows,

Table 1. Simulation parameters

| Channel Paths/CP Length | 5 / 8 |
|---|---|
| Sub-carriers | 32 |
| Length of pilot signals | 64 |
| Observed OFDM symbols | 64 |
| Equalizer | ZF |

We first consider only the main channel system case, system performance of the blind subspace (SS) channel estimation method compare with channel estimation add-on transmitting pilot signals assisted. Since the subspace estimation method based only on the received data signals, we can see a notable shortage in the BER (bit error rate) compared with the pilot signals assisted subspace estimation method. In Figure 12.

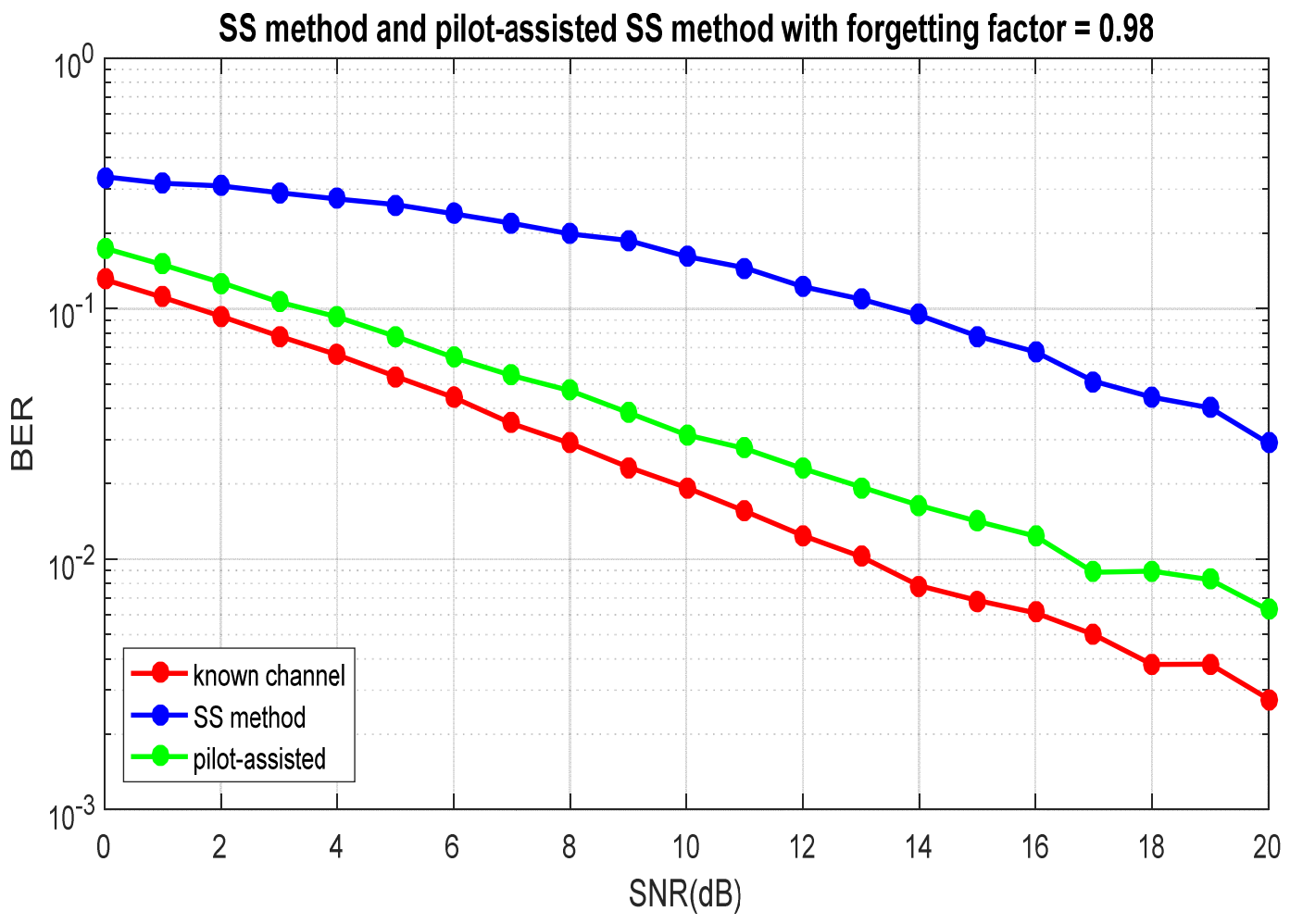

Figure 12. Main channel system estimation

Simulation results are mainly based on the information of nearby channels, the forgetting factor is set to 0.98. The BER decreases with the increase of the correlation coefficient of the two channels.

When the correlation coefficient is 0.98, system performance is quite satisfactory. The channel information-assisted subspace estimation method of the nearby channel is applicable. When the nearby channel correlation coefficients are 0.68 and 0.11 with 0.98 forgetting value, the effect form nearby channel information is reduced as shown in Figure 13. We then drop the forgetting factor value to 0.7 to decrease the impact of unreliable information from the nearby channel. Shown in Figure 14, the appropriate adjustment actually improve the estimation error caused by the not so related initial nearby channel information. In the process of adjusting the forgetting factor, it should be noticed that when the forgetting factor is less than a certain value, the rank of the received signal autocorrelation matrix will be reduced. When the rank is less than the number of symbols received, the subspace estimation method will no longer be valid.

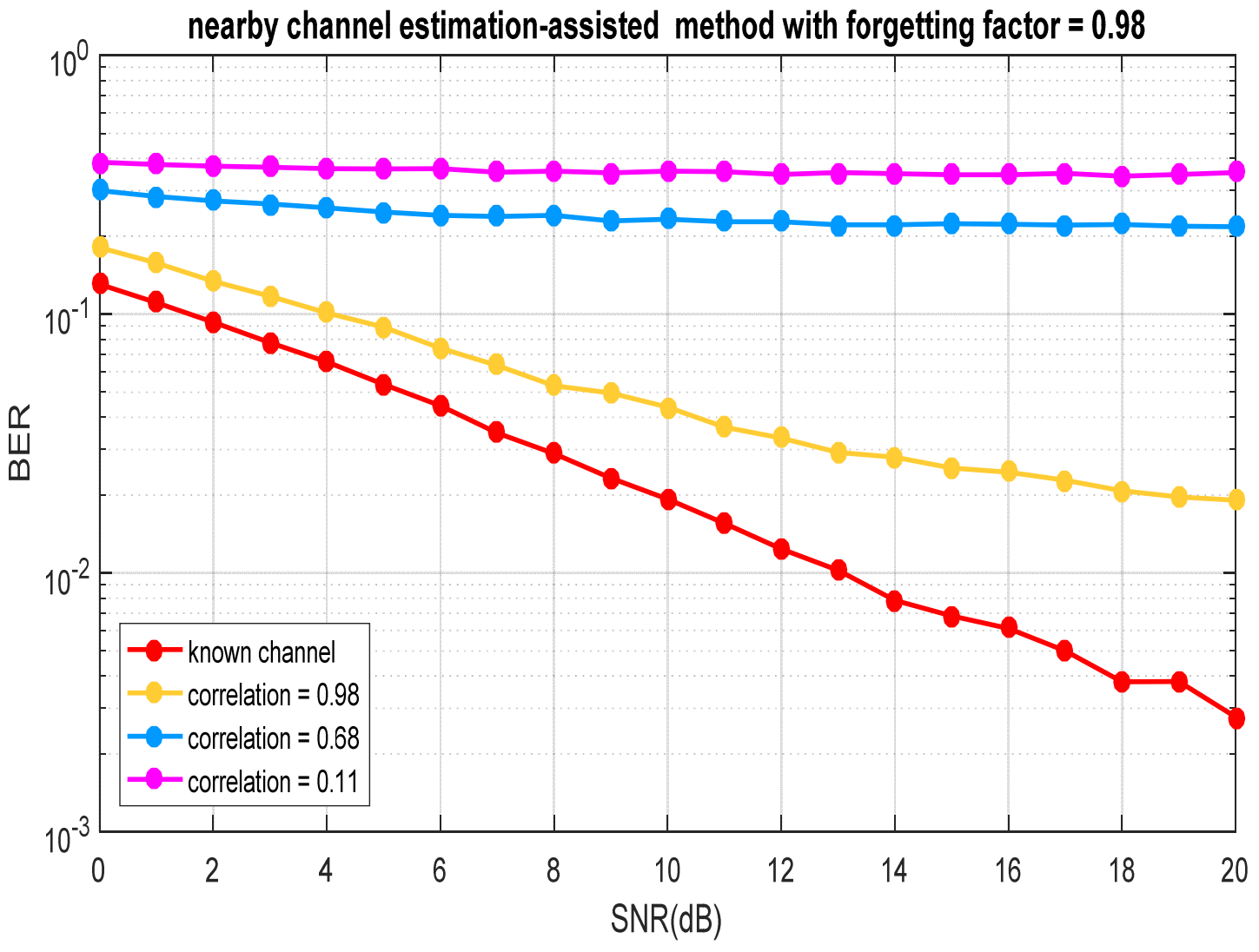

Figure 13. System with Nearby Channel Estimation

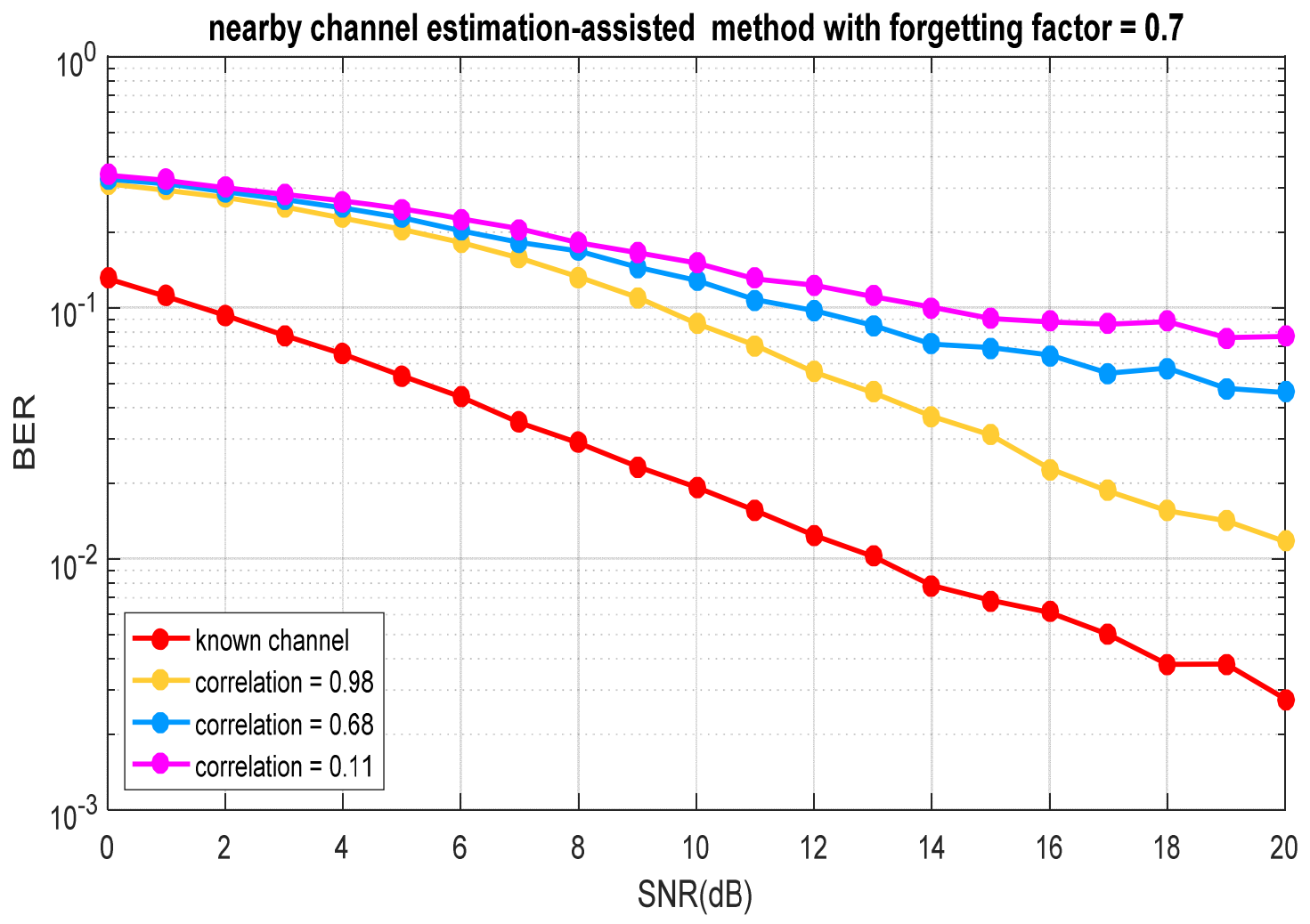

Figure 14. System with reduced Forgetting Factor

## CONCLUSIONS

This study aims to explore the impact of nearby channel estimation information on blind channel estimation. Since the blind estimation method need no pilot signals transmitting to save bandwidth requirement, we can obtain reliable initial channel information from nearby channel with pilot signals channel estimation. Simulation results show that the high correlation coefficient of the two nearby channels, the forgetting factor can be increased, so that the weight of the initial state of the matrix is increased, and a more accurate channel result is estimated. If the correlation coefficient is lowered, the forgetting factor can be appropriately adjusted to rely on the receive data signal more in the main channel to avoid high BER.

## ACKNOWLEDGMENTS

This work has been partially supported by the Ministry of Science and Technology, the Republic of China under grant number MOST 107-2119-M-019-004.